\documentclass[twocolumn,amsmath,prd]{revtex4}
\include{psfig}

\def\ga{\gamma}
\def\de{\delta}

\def\et{\eta}
\def\th{\theta}

\def\ta{\tau}

\def\vp{\varphi}
\def\ch{\chi}

\def\om{\omega}

\def\De{\Delta}
\def\Th{\Theta}

\def\Ph{\Phi}

\def\mn{{\mu\nu}}

\def\cl{{\cal L}}

\def\fr#1#2{{{#1} \over {#2}}}
\def\frac#1#2{\textstyle{{{#1} \over {#2}}}}
\def\pt#1{\phantom{#1}}

\def\vev#1{\langle {#1}\rangle}

\def\half{{\textstyle{1\over 2}}}
\def\lsim{\mathrel{\rlap{\lower4pt\hbox{\hskip1pt$\sim$}}
    \raise1pt\hbox{$<$}}}
\def\gsim{\mathrel{\rlap{\lower4pt\hbox{\hskip1pt$\sim$}}
    \raise1pt\hbox{$>$}}}

\def\etal {{\it et al.}}
\newcommand{\beq}{\begin{equation}}
\newcommand{\eeq}{\end{equation}}
\newcommand{\bea}{\begin{eqnarray}}
\newcommand{\eea}{\end{eqnarray}}
\newcommand{\bse}{\begin{subequations}}
\newcommand{\ese}{\end{subequations}}
\newcommand{\rf}[1]{(\ref{#1})}

\def\to{\rightarrow}

\def\mix{\leftrightarrow}
\def\nub{\bar\nu}
\def\vp{\vec p}

\def\heff{h_{\rm eff}}

\def\ring#1{{\mathaccent'27 #1}}
\def\cri{\ring{c}}

\def\a3em{\check{a}}
\def\cee{\cri}
\def\Dtm{\De m^2_{\Th}}
\def\Dtmz{\De m^2_{0^\circ}}

\def\lrDmu{\stackrel{\leftrightarrow}{D_\mu}}
\def\lrDnu{\stackrel{\leftrightarrow}{D^\nu}}

\begin{document}
\title{Lorentz and CPT violation in the neutrino sector}
\author{V.\ Alan Kosteleck\'y and Matthew Mewes}
\affiliation{Physics Department, Indiana University, 
         Bloomington, IN 47405, U.S.A.}
\date{IUHET 456, July 2003; 
accepted as a Rapid Communication, Phys.\ Rev.\ D} 

\begin{abstract}
We consider neutrino oscillations 
in the minimal Standard-Model Extension
describing general Lorentz and CPT violation.
Among the models without neutrino mass differences
is one with two degrees of freedom that reproduces 
most major observed features of neutrino behavior. 

\end{abstract}
\maketitle

%\pacs{11.30.Cp, 14.60.Pq}

Quantum physics and gravity 
are believed to combine at the Planck scale,
$m_P \simeq 10^{19}$ GeV.
Experimentation at this high energy is impractical,
but existing technology could detect 
suppressed effects from the Planck scale,
such as violations of relativity 
through Lorentz or CPT breaking 
\cite{cpt01}.
At experimentally accessible energies,
signals for Lorentz and CPT violation
are described by the Standard-Model Extension (SME)
\cite{ck},
an effective quantum field theory
based on the Standard Model of particle physics.
The SME incorporates general coordinate-independent
Lorentz violation.

The character of the many experiments
designed to study neutrino oscillations
\cite{pdg}
makes them well suited for tests of Lorentz and CPT symmetry.
The effects of Lorentz violation 
on propagation in the vacuum
can become more pronounced for light particles, 
and so small effects may become observable for large baselines.
Applying this idea to photons
has led to the best current sensitivity 
on any type of relativity violation
\cite{photonexpt}.

In this work,
we study the general neutrino theory given  
by the minimal renormalizable SME
\cite{ck}.
In this setup, 
as in the usual minimal Standard Model, 
SU(3)$\times$SU(2)$\times$U(1) symmetry is preserved,
the right-handed neutrino fields decouple and so are unobservable,
and there are no neutrino mass differences.
The neutrino behavior is contained in the terms
\bea
\cl &\supset&
\half i \overline{L}_a \ga^{\mu} \lrDmu L_a
- (a_L)_{\mu ab} \overline{L}_a \ga^{\mu} L_b
\nonumber\\
&&
\qquad\qquad
+ \half i (c_L)_{\mu\nu ab} \overline{L}_a \ga^{\mu} \lrDnu L_b
\label{SME}
\quad ,
\eea
where the first term is the usual Standard-Model kinetic term 
for the left-handed doublets $L_a$,
with index $a$ ranging over the three generations
$e$, $\mu$, $\ta$. 
The coefficients for Lorentz violation are
$(a_L)_{\mu ab}$,
which has mass dimension one 
and controls the CPT violation,
and $(c_L)_{\mu\nu ab}$,
which is dimensionless.
It is attractive to view these coefficients
as arising from spontaneous violation
in a more fundamental theory
\cite{kps},
but other origins are possible
\cite{cpt01}.

The Lorentz-violating terms in Eq.\ \rf{SME}
modify both interactions and propagation of neutrinos.
Any interaction effects are expected to be tiny 
and well beyond existing sensitivities.
In contrast, 
propagation effects can be substantial 
if the neutrinos travel large distances.
The time evolution of neutrino states
is controlled as usual by the effective hamiltonian
$(\heff)_{ab}$
extracted from Eq.\ \rf{SME}. 
The construction of $(\heff)_{ab}$
is complicated by the unconventional time-derivative term
but can be performed
following the procedure in Ref.\ \cite{kle}.
We find
\beq
(\heff)_{ab}=
|\vp| \de_{ab}
+ \fr{1}{|\vp|}
[ (a_L)^\mu p_\mu 
-(c_L)^\mn p_\mu p_\nu ]_{ab} .
\label{heff}
\eeq
To leading order,
the 4-momentum $p_\mu$ is $p_\mu=(|\vp|;-\vp)$.

The analysis of neutrino mixing proceeds along the usual lines.
We diagonalize $(\heff)_{ab}$
with a $3\times 3$ unitary matrix $U_{\rm eff}$,
$\heff =  U_{\rm eff}^\dag E_{\rm eff} U_{\rm eff}$,
where $E_{\rm eff}$ is a $3\times 3$ diagonal matrix.
There are therefore two energy-dependent eigenvalue differences
and hence two independent oscillation lengths, as usual. 
The time evolution operator is 
$S_{\nu_a\nu_b}(t)
= (U_{\rm eff}^\dag e^{-iE_{\rm eff}t} U_{\rm eff})_{ab}$,
and the probability for a neutrino of type $b$
to oscillate into a neutrino of type $a$ in time $t$ is 
$P_{\nu_b\to\nu_a}(t)=|S_{\nu_a\nu_b}(t)|^2$.

The CPT-conjugate hamiltonian $h^{\rm CPT}_{\rm eff}$
is obtained by changing the sign of $a_L$.
Under CPT, 
the transition amplitudes transform as 
$S_{\nu_a\nu_b}(t)
\leftrightarrow
S^*_{\nub_a\nub_b}(-t)$,
so CPT invariance implies
$P_{\nu_b\to\nu_a}(t) = P_{\nub_a\to\nub_b}(t)$.
Note that the converse is false in general
\cite{kmnu}.
For instance,
the model described below violates CPT 
but satisfies the equality.

Since oscillations are insensitive
to terms proportional to the identity,
each coefficient for Lorentz violation introduces 
two independent eigenvalue differences,
three mixing angles, and three phases.
The minimal SME (without neutrino masses)
therefore contains a maximum of 160 
gauge-invariant degrees of freedom
describing neutrino oscillations
\cite{fn1}.
Of these, 16 are rotationally invariant.
The existing literature  
concerns almost exclusively the rotationally invariant case
\cite{fc1,fc2,fc3,fc4},
usually with either $a_L$ or $c_L$ neglected
and in a two-generation model with nonzero neutrino masses.
A wealth of effects in the general case remains to be explored.

The presence of Lorentz violation introduces some novel features
not present in the usual massive-neutrino case.
One is an unusual energy dependence,
which can be traced to the dimensionality
of the coefficients for Lorentz violation. 
In the conventional case with mass-squared differences $\De m^2$,
neutrino oscillations are controlled by the dimensionless combination
$\De m^2L/E$
involving baseline distance $L$ and energy $E$.
In contrast,
Eq.\ \rf{heff} shows that oscillations due to
coefficients of type $a_L$ and $c_L$
are controlled by the dimensionless combinations
$a_L L$ and $c_L LE$, respectively.

Another unconventional feature is direction-dependent dynamics,
which is a consequence of rotational-symmetry violation.
For terrestrial experiments,
the direction dependence introduces sidereal variations 
in various observables
at multiples of the Earth's sidereal frequency
$\om_\oplus\simeq 2\pi/$(23 h 56 m).
For solar-neutrino experiments,
it may yield annual variations
because the propagation direction differs 
as the Earth orbits the Sun.
Both types of variations offer a unique signal 
of Lorentz violation
with interesting attainable sensitivities.
For solar neutrinos $LE\simeq 10^{25}$,
so a detailed analysis of existing data 
along the lines of Refs.\ \cite{sk}
might achieve sensitivities
as low as $10^{-28}$ GeV on $a_L$ and $10^{-26}$ on $c_L$
in certain models with Lorentz violation.
These sensitivities would be comparable to the best existing ones 
in other sectors of the SME
\cite{photonexpt,cavexpt,ccexpt,eexpt1,eexpt2,muons,hadronexpt,fn3}.

The coefficients for Lorentz violation can also 
lead to novel resonances, 
in analogy to the MSW resonance
\cite{msw}.
Unlike the usual case,
however,
these Lorentz-violating resonances can occur also in the vacuum
and may have directional dependence
\cite{fn4}.
Note that conventional matter effects
can readily be handled within our formalism \rf{heff}
by adding the effective contributions
$(a_{L, \rm eff})^0_{ee}=G_F(2n_e-n_n)/\sqrt{2}$
and
$(a_{L, \rm eff})^0_{\mu\mu}
=(a_{L, \rm eff})^0_{\ta\ta}=-G_Fn_n/\sqrt{2}$,
where $n_e$ and $n_n$ are the number
densities of electrons and neutrons.
The contributions to $\heff$ from matter range from 
about $10^{-20}$ GeV to $10^{-25}$ GeV.
This range is within the region expected 
for Planck-scale Lorentz violation,
so matter effects can play a crucial role in the analysis.

An interesting question is whether the introduction of Lorentz violation 
may help explain the small LSND excess of $\nub_e$
\cite{lsnd}.
Usually,
two mass-squared differences are invoked to explain the observations
in solar and atmospheric neutrinos,
but LSND lies well outside the region of limiting
sensitivity to these effects.
Possible solutions to this puzzle may arise
from the unusual energy and directional dependences 
of Lorentz violation.
An explanation of LSND requires 
a mass-squared difference of about
$10^{-19}$ GeV$^2 = 10^{-1}$ eV$^2$,
an $a_L$ of about $10^{-18}$ GeV,
or a $c_L$ of about $10^{-17}$.
Any of these would affect other experiments 
to some degree, 
including the MiniBooNE experiment
\cite{miniboone} 
designed to test the LSND result.

To illustrate some of the possible behavior
allowed by the SME,
we consider a two-coefficient three-generation case
without any mass-squared differences,
but incorporating an isotropic $c_L$ with nonzero element 
$\fr43(c_L)^{TT}_{ee}
\equiv 2\cee$
and an anisotropic $a_L$ 
with degenerate nonzero real elements 
$(a_L)^Z_{e\mu}=(a_L)^Z_{e\ta}\equiv\a3em/\sqrt{2}$.
The coefficients are understood to be specified in 
the conventional Sun-centered celestial equatorial frame $(T,X,Y,Z)$,
which has $Z$ axis along the Earth rotation axis
and $X$ axis towards the vernal equinox
\cite{kmphot}.
In what follows,
we show that this simple model,
which we call the `bicycle' model, 
suffices to reproduce the major features 
of the known neutrino behavior 
other than the LSND anomaly,
despite having only two degrees of freedom
rather than the four degrees of freedom
used in the standard description with mass. 

Diagonalizing the hamiltonian for the model yields 
$$P_{\nu_e\to\nu_e}=
1-4\sin^2\th\cos^2\th\sin^2(\De_{31}L/2) ,
$$
$$
P_{\nu_e\mix\nu_\mu}=P_{\nu_e\mix\nu_\ta}
=2\sin^2\th\cos^2\th\sin^2(\De_{31}L/2) ,
$$
\bea
P_{\nu_\mu\to\nu_\mu}
&=& P_{\nu_\ta\to\nu_\ta}
=1-\sin^2\th\sin^2(\De_{21}L/2) \nonumber\\
&&
\pt{= P_{\nu_\ta\to\nu_\ta}}
-\sin^2\th\cos^2\th\sin^2(\De_{31}L/2) \nonumber\\
&&
\pt{= P_{\nu_\ta\to\nu_\ta}}
-\cos^2\th\sin^2(\De_{32}L/2) , \nonumber \\
P_{\nu_\mu\mix\nu_\ta}&=&
\sin^2\th\sin^2(\De_{21}L/2) \nonumber\\
&&-\sin^2\th\cos^2\th\sin^2(\De_{31}L/2) \nonumber\\
&&+\cos^2\th\sin^2(\De_{32}L/2) ,
\eea
where
\bea
\De_{21}&=&\sqrt{(\cee E)^2+(\a3em\cos\Th)^2}+\cee E , 
\nonumber \\
\De_{31}&=&2\sqrt{(\cee E)^2+(\a3em\cos\Th)^2} ,
\nonumber \\
\De_{32}&=&\sqrt{(\cee E)^2+(\a3em\cos\Th)^2}-\cee E ,
\nonumber \\
\sin^2\th&=&\half [1-{\cee E}/
{\sqrt{(\cee E)^2+(\a3em\cos\Th)^2}}] ,
\label{sinth}
\eea
and where we define the propagation direction by the unit vector 
$\hat p=(\sin\Th\cos\Ph,\sin\Th\sin\Ph,\cos\Th)$
in polar coordinates in the standard Sun-centered frame.
These probabilities also hold
for antineutrinos.

The qualitative features of the model
can be understood as follows.
At low energies, $\a3em$ causes oscillation 
of $\nu_e$ into an equal mixture of $\nu_\mu$ and $\nu_\ta$.
At high energies, 
$\cee$ dominates and prevents $\nu_e$ mixing.
For definiteness, we take $\cee >0$.
At energies well above the critical energy $E_0=|\a3em |/\cee$,
$\sin^2\th$ vanishes
and the probabilities reduce to a maximal-mixing two-generation 
$\nu_\mu\mix\nu_\ta$ case with transition probability
$P_{\nu_\mu\mix\nu_\ta}\simeq\sin^2(\De_{32}L/2)$,
$\De_{32}\simeq\a3em^2\cos^2\Th/2\cee E$. 
The energy dependence in this limit is therefore
that of a conventional mass-squared difference 
of $\Dtm\equiv\a3em^2\cos^2\Th/\cee$. 
This pseudomass appears because 
the hamiltonian contains one large element at high energies,
triggering a Lorentz-violating seesaw.
Other models using combinations
of mass and coefficients for Lorentz violation 
can be constructed to yield various exotic 
$E^n$ dependences at particular energy scales.
Note that the high-energy pseudomass 
and hence neutrino oscillations
depend on the declination $\Th$ of the propagation.
High-energy neutrinos propagating
parallel to celestial north or south
experience the maximum pseudomass $\Dtmz = \a3em^2/\cee$,
while others see a reduced value $\Dtm=\Dtmz\cos^2\Th$.
For propagation in the equatorial plane, 
all off-diagonal terms in $\heff$ vanish
and there is no oscillation.

\begin{figure}
\centerline{\psfig{figure=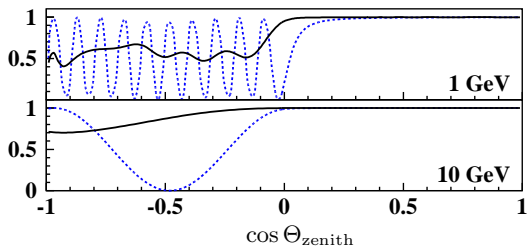,width=0.8\hsize}}
\centerline{\hspace*{25pt}$\cos\Th_{\rm zenith}$}
\caption{\label{caatmos}
$P_{\nu_\mu\to\nu_\mu}$
averaged over azimuthal angle 
for the bicycle model (solid)
and for a conventional case with mass (dotted).}
\end{figure}

\begin{figure}
\centerline{\psfig{figure=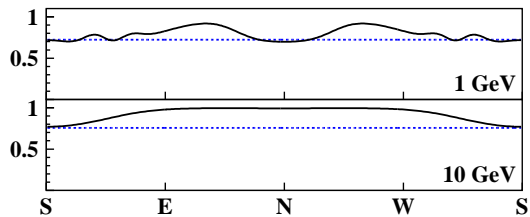,width=0.8\hsize}}
\caption{\label{caatmos2}
$P_{\nu_\mu\to\nu_\mu}$
averaged over zenith angle
for the bicycle model (solid)
and for a conventional case with mass (dotted).}
\end{figure}

The features of atmospheric oscillations in the model
are compatible with published observations.
For definiteness,
we take $\Dtmz$ near the accepted range 
required in the usual analysis
and $E_0$ below the relevant energies:
$\Dtmz=10^{-3}$ eV$^2$ and $E_0=0.1$ GeV
($\cee=10^{-19}$, $\a3em=10^{-20}$ GeV).
High-energy atmospheric neutrinos then exhibit 
the usual energy dependence,
despite having zero mass differences.
The zenith-angle dependence of the
probability $P_{\nu_\mu\to\nu_\mu}$
averaged over the azimuthal angle 
also is comparable 
within existing experimental resolution
to a conventional maximal-mixing case
with two generations and a mass-squared difference
$\De m^2=2\times10^{-3}$ eV$^2$,
as is shown in Fig.\ \ref{caatmos}
for latitude $\ch\simeq36^\circ$.
However,
the model predicts significant
\it azimuthal \rm 
dependence for atmospheric neutrinos,
which is a signal for Lorentz violation.
For example, 
consider neutrinos propagating in the horizontal plane
of the detector.
Neutrinos originating from the east or west have $\cos\Th=0$,
$\Dtm=0$, and hence no oscillations.
In contrast, 
those entering the detector from the north or south
experience a pseudomass of $\Dtm=\Dtmz\cos^2\ch$.
Figure \ref{caatmos2} shows the survival probability
averaged over zenith angle as a function of azimuthal angle.
Although this model predicts no east-west asymmetry 
beyond the usual case,
north-east or north-south asymmetries appear.
Similar `compass' asymmetries are typical 
in all direction-dependent models.

\begin{figure}
\centerline{\psfig{figure=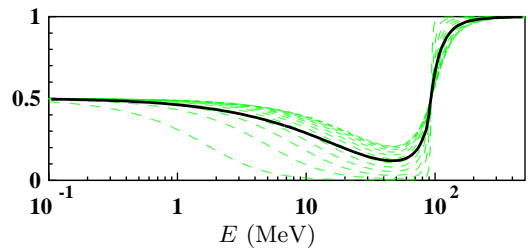,width=0.8\hsize}}
\centerline{$E$ (MeV)}
\caption{\label{casol}
$(P_{\nu_e\to\nu_e})_{\rm adiabatic}$
averaged over one year (solid) and
at intervals between an equinox and a solstice (dashed).}
\end{figure}

The basic features of solar-neutrino oscillations predicted
by the model are also compatible with observation.
Observed solar neutrinos propagate in the Earth's orbital plane,
which lies at an angle $\et\simeq23^\circ$ 
relative to the equatorial plane.
The value of $\cos^2\Th$
therefore varies from zero at the two equinoxes
to its maximum of $\sin^223^\circ$
at the two solstices.
Assuming adiabatic propagation in the Sun,
the average $\nu_e$ survival probability is 
\beq
(P_{\nu_e\to\nu_e})_{\rm adiabatic}=
\sin^2\th\sin^2\th_0+\cos^2\th\cos^2\th_0 ,
\label{camsw}
\eeq
where $\th_0$ is the mixing angle at the core,
given by replacing
$-\cee E$ with $-\cee E + G_Fn_e/\sqrt{2}$
in Eq.\ \rf{sinth}.
Figure \ref{casol} shows
the adiabatic probability as a function of energy
averaged over one year.
The predicted neutrino flux 
is half the expected value at low energies
and decreases at higher energies,
consistent with existing data.
Also shown is the adiabatic probability at approximately
weekly intervals between an equinox and a solstice.
Over much of the year,
it remains near the average.
There is a strong reduction near each equinox,
but the adiabatic approximation fails there 
because oscillations cease,
and so the true survival probability peaks sharply to unity.
The combination of effects produces ripples in the binned flux
near the equinoxes,
which might be detected in detailed experimental analyses
of existing or future data.

\begin{figure}
\centerline{\psfig{figure=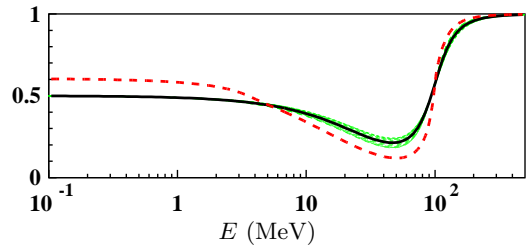,width=0.8\hsize}}
\centerline{$E$ (MeV)}
\caption{\label{casolmod}
$(P_{\nu_e\to\nu_e})_{\rm adiabatic}$
for some modified models.
}
\end{figure}

Although detection of the semiannual variation
would represent a definite positive signal for Lorentz violation,
its absence cannot serve to eliminate this type of model. 
Simple modifications of the model exist 
that exhibit similar overall behavior 
for solar and atmospheric neutrinos
but have only a small semiannual variation.
As an illustration,
consider the replacement of the coefficient $(a_L)^Z_{e\mu}$
with a coefficient $(a_L)^T_{e\mu}$ 
of half the size.
This has the effect of replacing 
the solid and dashed curves 
of Fig.\ \ref{casol} 
with those shown in Fig.\ \ref{casolmod}.
The semiannual variations in this type of model lie below 
existing statistical sensitivities.
Replacing also $(a_L)^Z_{\mu\ta}$ with $(a_L)^T_{\mu\ta}$
is another option, 
which removes all orientation dependence in the model. 
Another example of a small modification 
is a 10\% admixture of $(a_L)^T_{ee}$,
which  raises the survival probability of 0.5 at low energies 
to about 0.6 without appreciably affecting other results.
The ensuing survival probability in the adiabatic approximation
is shown as the dotted line in Fig.\ \ref{casolmod}. 
Other more complicated modifications that could be countenanced
but that nonetheless retain the flavor of the simple model 
include allowing dependence on directions other than $Z$,
or even introducing arbitrary coefficients  
$(a_L)^\mu_{ee}$, $(a_L)^\mu_{e\mu}$, $(a_L)^\mu_{e\ta}$, 
and $(c_L)^{\mu\nu}_{ee}$, 
which yields a model with 21 degrees of freedom.
More general possibilities also exist
\cite{kmnu}.
We conclude that positive signals for Lorentz violation
could be obtained by detailed fitting of existing experimental data,
but that it is challenging and perhaps even impossible at present
to exclude the possibility 
that the observed neutrino oscillations are due to
Lorentz and CPT violation rather than to mass differences. 

The observations from long-baseline experiments
are also compatible with the oscillation lengths in the 
simple two-coefficient model.
For example,
the oscillation length $2\pi/\De_{31}$ 
controls $\nub_e$ survival
and is short enough to affect KamLAND
\cite{kamland}.
An analysis incorporating the relative locations of the detector 
and the individual reactors would be of definite interest
but lies outside our scope.
Note, however, that the average
$\nub_e$ survival probability is 
$\vev{P_{\nub_e\to\nub_e}}
=1-2\sin^2\th\cos^2\th \ge 1/2$.
A complete analysis is therefore likely to yield 
a reduced flux somewhat more than half the expected flux,
in agreement with current data.

The new class of long-baseline 
accelerator-based experiments
\cite{lbex},
planning searches for oscillations in $\nu_\mu$
at GeV energy scales 
and distances of hundreds of kilometers,
will be sensitive to sidereal variations.
The model predicts $\nu_\mu\mix\nu_\ta$ mixing 
with an experiment-dependent pseudomass 
$\Dtm = \Dtmz\cos^2\Th$
because their beamlines are in different directions
and so involve a different propagation angle $\Th$.
The energy dependence and transitions 
will be similar to the usual mass case.

Although the simple bicycle model
reproduces most major features of observed neutrino behavior, 
it incorporates only a tiny fraction 
of the many possibilities allowed in the SME.
More complexity could be introduced in performing 
a detailed fit to all existing data.
Nonetheless,
the model serves to illustrate a few key phenomena 
introduced by Lorentz violation.
It also shows that 
the presence of Planck-scale Lorentz and CPT violation 
in nature could well first be revealed by a definitive signal
in neutrino oscillations.

This work was supported in part 
by D.O.E.\ grant DE-FG02-91ER40661
and by N.A.S.A.\ grants NAG8-1770 and NAG3-2194.


\begin{thebibliography}{99}
\bibitem{fn1}
Some of these are unobservable 
because they can be removed by suitable field redefinitions.
\bibitem{fn4}
An explicit example in a rotationally invariant model
with a timelike coefficient $(a_L)^T$ is studied 
in Ref.\ \cite{fc2}.
\end{thebibliography}
\end{document}